\documentclass{article}
\usepackage{amssymb,amsmath,amsthm,graphicx,epsfig,latexsym}

\def\a{{\alpha}}
\def\b{{\beta}}
\def\g{{\gamma}}

\def\t{{\tau}}

\def\L{{\Lambda}  }

\def\R{{\mathbb{R}}}
\def\Z{{\mathbb{Z}}}
\def\Z2{\mathbb{Z}_2}

\def\T{\mathbb{T}  }
\def\V{\mathcal{V} }

\def\EA{\Sigma }      
\def\EAL{{\EA}_{\L}  }
\def\EV{{2^{\V} }  }          
\def\EM{{2^{\MCE} }  }
\def\EMS{{2^{\M} }  }

\def\muL{\mu_{\L}   }
\def\P{\mathbb{P}  }
\def\f{\phi}            
\def\fp{f_{\phi}  }
\def\CE{{\EA^{\diamond}}}        
\def\MCE{{\EA^{*}}  }        
\def\S{\Omega }     
\def\H{(\S,\EA,\mu) }     
\def\HL{(\L,\EAL,\muL) }
\def\HP{(\S,\EA,\P) }

\def\vP{{\triangle P} }
\def\vM{\triangle\MCE }
\def\vsupp{\triangle supp }

\def\sP{\varpi }
\def\sO{\mathcal{O}  }

\def\C{{\cal{C}}(\Omega,\EA,\mu)} 
\def\M{{\cal{M}}(\Omega,\EA,\mu)} 
\def\F{{\cal{F}} }
\def\Fs{ {\cal{F}}_{*} }
\def\U{ \cal{U}  }

\def\leqs{\leq_{\subseteq}  }
\def\wedges{\wedge_{\subseteq}  }
\def\vees{\vee_{\subseteq}  }

\def\leqt{\leq_{\t}  }
\def\wedget{\wedge_{\t}  }
\def\veet{\vee_{\t}  }

\newtheorem{theorem}{Theorem}

\newtheorem{lemma}{Lemma}
\newtheorem{corollary}{Corollary}

\newtheorem{definition}{Definition}

\theoremstyle{remark}
\newtheorem{example}{Example}[section]

\newcommand\beq{\begin{equation}}
\newcommand\eeq{\end{equation}}
\newcommand\bea{\begin{eqnarray}}
\newcommand\eea {\end{eqnarray}}
\newcommand\ba{\begin{array}}
\newcommand\ea {\end{array}}

\newcommand\bd{\begin{description}}
\newcommand\ed {\end{description}}
\newcommand\ben{\begin{enumerate}}
\newcommand\een{\end{enumerate}}

\newcommand\bD{\begin{definition} }
\newcommand\eD{\end{definition} }
\newcommand\bE{\begin{example} }
\newcommand\eE{\end{example} }
\newcommand\bL{\begin{lemma} }
\newcommand\eL{\end{lemma} }
\newcommand\bT{\begin{theorem} }
\newcommand\eT{\end{theorem} }

\newcommand\bC{\begin{corollary} }
\newcommand\eC{\end{corollary} }

\def\nn{{\nonumber} }

\begin{document}

\vfill


\vfill
\vfill

\begin{center}
   \baselineskip=16pt
   \begin{LARGE}
      \textsl{Coevents as Beables}
   \end{LARGE}
   \vskip 2cm
      Yousef Ghazi-Tabatabai
   \vskip .6cm
   \begin{small}
      \textit{yousef.ghazi05@imperial.ac.uk}
        \end{small}
\end{center}

\vskip 1cm

\begin{small}
\begin{center}
   \textbf{Abstract}
\end{center}
Sorkin's \emph{coevents} \cite{Sorkin:coeventsI} can be thought of as the `beables' of a quantum histories theory \cite{Sorkin:coeventsII}; in this paper we study the `logical' implications of taking this claim at face value, constructing a propositional lattice for the space of coevents applicable to a given histories theory and comparing it to the more traditional propositional lattice of events. In particular we focus on multiplicative coevents, and find that the precise nature of their anhomomorphism leads to a particularly simple relationship between the two propositional lattices. Finally, we notice that our constructions contain elements intuitively similar to topos nations of truth values and use this to suggest a means of applying Isham's topos theoretic constructions \cite{Isham:1997,Isham:toposI} to multiplicative coevents.
\end{small}
\vskip 0.5cm

\section{Introduction}\label{sec:introduction}

\subsection{Opening Remarks}\label{subsec:openning remmarks}

\emph{Quantum measure theory} \cite{Sorkin:1994dt,Sorkin:1995nj} rephrases the \emph{histories approach} to quantum mechanics \cite{Hartle:1992as,Gell:1990,Omnes:1988ek,Omnes:1988ek} as a generalization of a classical probability theory, with a sample space consisting of the spacetime paths introduced by Dirac \cite{Dirac:1933} and Feynman \cite{Feynman:1948,Feynman:1965}. A naive application of the classical `one history is real' interpretation is obstructed by the Kochen-Specker theorem \cite{Yousef:KSP,Kochen:1967}, leading to the development of various alternative interpretations.

Perhaps the most prominent of these alternatives is the \emph{consistent histories} interpretation \cite{Griffiths:1984rx,Griffiths:1993,Griffiths:1996,Griffiths:1998}, more recently topos based ideas have also been introduced \cite{ Isham:1997,Isham:toposI,Isham:toposII,Isham:toposIII,Isham:toposIV,Butterfield:2003,Flori:2008fb,Isham:1999kb}. Sorkin has proposed the \emph{coevent interpretation} \cite{Sorkin:coeventsI,Sorkin:coeventsII,Sorkin:2010,Yousef:CoeventDynamics,Yousef:KSP,Yousef:thesis,Gudder:2009a,Gudder:2009b,Gudder:2010}, focusing on truth valuation maps (\emph{coevents}) $\phi:\EA\rightarrow\Z2$ from the event algebra of a histories theory to the truth value space $\Z2$. In this context the classical `one history is real' can be rephrased as `$\phi$ is a homomorphism', the generalisation of which leads to the `anhomomorphic logic' of `non-classical coevents'.

\subsection{Goals and Outline of this Paper}\label{subsec:outline of paper}

In \cite{Sorkin:coeventsII} Sorkin has proposed two ways of understanding the ontology of the coevent interpretation, firstly that histories are the beables with coevents describing a `non-classical', `anhomomorphic' logical framework. Alternatively we might regard the coevents themselves as the beables, in which case we are able to use a `standard' or `classical' logical framework. In this paper we take the first steps toward a rigorous implementation of the second approach, examining `logical' implications of taking coevents seriously as beables. Following from \cite{Yousef:CoeventDynamics} in which we examine the construction of dynamics for coevents, we now turn to the `logical structure' of propositions concerning coevents and their relationship with propositions concerning histories.

Our use of the phrase `logical structure' itself requires some interpretation; we will understand it to refer to the lattice structure of a set of propositions. In the case of a probability or histories theory we will expect our propositional lattice to be the event algebra ordered by inclusion, with the lattice structure defined by the order relation. We will formalise the notion of the `threefold character' of logic discussed in \cite{Sorkin:coeventsII} by extending the idea of a logical structure to a \emph{logical framework}; associating with our propositional lattice $\EA$ a truth value space $\T$ and a set of `allowed' truth valuation functions $\V$ from the lattice to this space\footnote{Note that since domain and range are inherent in the definition of a function the propositional lattice and truth valuation space are redundant in a logical framework. We will however continue to explicitly include them for clarity.} (which for our purposes will always be $\Z2$) to yield a triple $(\EA,\V,\T)$. For example, the lattice structure of a histories event algebra together with $\Z2$ and a coevent scheme \cite{Yousef:thesis} constitutes a logical framework for a histories theory. In what follows we will construct logical frameworks for coevents themselves, and examine how these relate to the logical frameworks of the associated histories theories.

In section $1.3$ we review classical probability theories, rephrasing them in a manner which anticipates quantum measure theory and the coevent interpretation. In section $1.4$ we briefly review quantum measure theory and in $1.5$ we introduce coevents and the multiplicative scheme. In section $2$ we undertake our central task of constructing logical frameworks for coevents and examining their relationship with propositions concerning histories. In section $2.1$ we lay out our general construction, while in section $2.3$ we carefully apply it to multiplicative coevents. In section $3$ we note that a key piece of our construction is reminiscent of elements of topos theory, leading us to examine the applicability of such methods to the coevent interpretation and the analysis of section $2$. We conclude in section $4$.

\subsection{Classical Physics}\label{subsec:classical physics}

\subsubsection{Classical Dynamics: Probability Theories}\label{subsubsec:classical dynamics}

We begin by describing classical dynamics and ontology as a starting point from which to generalise to quantum theories. A classical probability theory is a triple $\HP$, where $\S$ is the \emph{sample space}, $\EA$ the \emph{event algebra} and $\P$ the \emph{(probability) measure}.

The sample space is in a sense our most basic object, and can for our purposes be thought of as the space of full specifications of the system under study; looking ahead we denote the elements of the sample space as \emph{histories}. For example if we are studying a single throw of a `fair coin' our sample space will be $\{h,t\}=\{heads,tails\}$, if we consider two throws we will have $\S=\{hh,ht,th,tt\}$. In what follows we will for simplicity assume that $\S$ is finite.

The event algebra $\EA\subset 2^{\S}$ can be thought of as the space of propositions we make concerning the system, we often take $\EA$ to be a sigma subalgebra of $2^{\EA}$ but when $\S$ is finite we can use the whole space $\EA=2^{\S}$. Set inclusion equips the event algebra with a natural partial order, $A\leq B \Leftrightarrow A\subset B$, which is easily seen to make $\EA$ a Boolean lattice where meet, join and complement are equivalent to the set relations intersection, union and complement respectively; this lattice structure can be thought of as the natural `logical structure' of our space of propositions. We will in what follows use lattice structure synonymously with `logical structure', thinking of the former as a concrete realisation of the latter. Notice that by writing $AB=A\cap B$ and $A+B=A\triangle B$ (where $\triangle$ denotes the symmetric difference) we can formulate $\EA$ as an algebra over $\Z2$.

Finally the dynamics are encoded in the probability measure $\P$, a positive real valued function on the event algebra with normalisation $\P(\S)=1$ obeying the \emph{Kolmogorov sun rule},
\beq\label{eq:Kolmogorov sun rule}
\P(A\sqcup B) = \P(A)+\P(B),
\eeq
for all disjoint $A,B\in\EA$, where $\sqcup$ denotes disjoint union. Notice that $\P$ encodes all dynamical information, including `initial conditions'.

\subsubsection{Classical Ontology: Truth Valuation Maps}\label{subsubsec:classical ontology}

The ontology of this classical theory can be summed up in the statement that a single element $r\in\S$ of the sample space is deemed to be `\emph{real}'. Events which contain this real history are said to be true while events which do not contain it are said to be false. Now in general we do not know which event `really happens', so we construct the space of potentially real events, which we call the \emph{ontology} of our theory; we will denote elements of this set as \emph{potential realities} or \emph{beables}. Note that our choice of ontology is affected by our knowledge concerning the system; if we know that the real event is $r$ then our ontology is simply $\{r\}$, if we can place no restriction on which event is real then our ontology will be the full sample space $\S$. In general the information we have about the system is represented by the dynamics (which as we have seen includes initial conditions), which in our classical probability theories means the probability measure $\P$. There is some discussion around the interpretation of probability and its relationship to ontology; we will simply use the most basic connection, the concept of \emph{preclusion} which requires events of measure zero to be `false'. From the above this immediately implies that no element of a measure zero set can be `real', so our ontology now becomes the set,
\beq
\{\g\in\S \ | \ \nexists A\ni \g \ \text{such that} \ \P(A)=0\}.
\eeq

We can now construct a `logical framework' for $\HP$; we have already identified the lattice $\EA$ as the `logical structure', we now use the ontology to provide our `allowed' truth values. We have seen that if the `real' history is $r\in\S$ then any event $A\in\EA$ containing $r$ is `true' and every event not containing $r$ is `false'. Identifying the `truth value space' $\{false,true\}$ with $\{0,1\}=\Z2$, this directly defines a truth valuation function which we will call $r^*$,
\bea
r^*:\EA&\rightarrow&\{false,true\}=\{0,1\}=\Z2 \nn \\
r^*(A) &=& \left\{\begin{array}{cc} 1 & r\in A \\ 0 & r\not\in A \\ \end{array}\right. \label{eq:classical coevent}
\eea
Now notice that $\Z2$ is a Boolean lattice (and thus an algebra over $\Z2$), it is easy to see that these `classical' truth valuation maps are lattice (and thus algebra) homomorphisms; in fact they are all the homomorphisms,
\beq
\{\g^* \ | \ \g\in\S\} = Hom(\EA,\Z2),
\eeq
where $Hom(A,B)$ is the space of homomorphisms from the lattice (algebra) $A$ to the lattice (algebra) $B$.

Finally, we can translate the concept of preclusion into our truth valuation perspective. The requirement that dynamically precluded events are also ontologically precluded sits naturally with an ontology of truth valuation functions, we simply require that for truth valuation map $r^*$ to be part of our ontology it must satisfy,
\beq
\P(A)=0\Rightarrow r^*(A)=0.
\eeq
We call such maps \emph{preclusive}, and will label the set of preclusive homomorphisms from the event algebra to $\Z2$ by ${\cal{C} }(\S,\EA,\P)$. We can then think of the triple $(\EA,{\cal{C} }(\S,\EA,\P),\Z2)$ as our logical framework.

\subsection{Quantum Dynamics}\label{subsec:quantum dynamics}

\subsubsection{Quantum Measures: Histories Theories}\label{subsubsec:histories theories}

We must generalise the above structure to accommodate quantum dynamics; we begin by generalising a classical probability theory to a \emph{histories theory} $\H$, where as before $\S$ is our sample space, or histories space, which we will assume to be finite, and the Boolean lattice $\EA=2^{\S}$ is our event algebra of propositions. We have however generalised the dynamics from a probability measure to a \emph{quantum measure} $\mu$ which does not necessarily obey the Kolmogorov sum rule but instead is constrained by the \emph{level 2 sum rule}. More formally we call $\mu:\EA\rightarrow\R$ a quantum measure if it obeys,
\bea
\mu(A)&\geq& 0 \nn \\
\mu(\S)&=&1 \nn \\
\mu(A\sqcup B\sqcup C) &=& \mu(A\sqcup B)+\mu(B\sqcup C)+\mu(C\sqcup A) \nn \\
&& -\mu(A)-\mu(B)-\mu(C). \label{eq:level 2 sum rule}
\eea

A quantum measure is similar in dynamical content to the more standard decoherence functional $D:\EA\times\EA\rightarrow {\mathbb{C}}$ from which it can be derived,
\beq
\mu(A)=D(A,A).
\eeq

We refer to a partition $\L$ of $\S$ as a \emph{coarse graining} (and $\S$ is called a \emph{fine graining} of $\L$); $\L$ generates a Boolean subalgebra $\EAL$ of $\EA$ to which we can restrict the measure leading to the \emph{coarse grained histories theory} $\HL$. In some cases the restriction of the measure to a particular subalgebra $\EAL$ might obey the Kolmogorov sum rule, so that the coarse grained theory $\HL$ is a classical probability theory. We call such subalgebras (and their associated partitions) \emph{dynamically classical} or \emph{decoherent}.

\subsection{Coevents}\label{subsec:valuations intro}

\subsubsection{Basic Definitions}\label{subsubsec:coevents}

It is not clear how we might naively apply the classical structures defined above to a general quantum histories theory; in particular there are several no-go theorems obstructing the use of a single history as a truth valuation map. Given a histories theory $\H$ we translate the concept of preclusion to $\mu(A)=0\Rightarrow\g^*(A)=0$ for $\g\in\S$, allowing us to define $\C$ as we did for classical probability theories. Then it is possible to find gedanken-experimentally realisable systems (based for example on the Kochen-Specker Theorem \cite{Yousef:KSP}) in which the sample space is covered by null sets, leading to $\C=\emptyset$.

The co-event approach generalises from classical homomorphic truth valuation maps to arbitrary maps from the event algebra to $\Z2$.
\begin{definition}
Let $\H$ be a histories theory, then we refer to a map,
\beq\nn
\phi:\EA\rightarrow\Z2,
\eeq
as a coevent; we denote the space of coevents by $\CE$. A coevent is preclusive if
\beq\nn
\mu(A)=0\Rightarrow\phi(A)=0 \ \ \forall A\in\EA.
\eeq
\end{definition}

Because our coevents can be anhomomorphic we have to pay close attention to difference between the `logic' (lattice structure) in the domains and the truth value space; as always we must be careful to distinguish both from our `meta-logic'\footnote{Our meta-logic includes the mathematical reasoning we use to discuss the logical structures under examination.}. For example, when discussing the multiplicative scheme, and using $\rightarrow$ to denote implication\footnote{We can define implication in a Boolean lattice in the standard fashion, $A\rightarrow B=\neg A\vee B$.} in $\EA$ and $\rightsquigarrow$ to denote implication in the truth value space ($\Z2$), we can write Modus Ponens as
\beq\label{eq:Modus Ponens}
(A\rightarrow B) \Rightarrow (\f(A)\rightsquigarrow\f(B)).
\eeq
To avoid confusion we will use the symbols $\curlywedge, \curlyvee, \rightharpoondown, \rightsquigarrow$ to denote meet, join, complement and implication in the truth value space to emphasise that these operations are distinct from those in the event algebra.

If a coevent is a homomorphism we say that it is \emph{classical}, thus $\C$ is the space of classical, preclusive coevents over the histories theory $\H$. As we have seen above, in a classical probability theory (when $\mu$ is decoherent on all of $\EA$) we consider $\C$ to be the set of `dynamically allowed' coevents. In a general histories theory we will take some subset $\V\subseteq\CE$ as the set of allowed coevents; various means have been suggested for choosing this subset. The choice of $\V$ is referred to as a \emph{coevent scheme}.

\subsubsection{The Multiplicative Scheme}\label{subsubsec:multiplicative scheme}

The current `standard approach' is known as the \emph{multiplicative scheme}, which places a sequence of restrictions on $\CE$. First, we say that a coevent is multiplicative if,
\beq
\f(A\wedge B)=\f(A)\curlywedge\f(B) \ \ \forall A,B\in\EA.
\eeq
Now regarding both $\EA$ and $\Z2$ as algebras over $\Z2$ we have $A\wedge B=AB$ and $\f(A)\curlywedge\f(B)=\f(A)\f(B)$, keading us to rephrase the above condition as,
\beq\label{eq:multiplicativity}
\f(AB)=\f(A)\f(B) \ \ \forall A,B\in\EA,
\eeq
which makes the choice of the term `multiplicative' more clear. We denote the space of multiplicative coevents by $\MCE$. In terms of the logical structure, multiplicative coevents preserve the `AND' but not the `OR' relation, in other words they preserve meets but not joins; we will return to this topic in later sections.

It is easy to see that the support of a multiplicative coevent is a filter \cite{Yousef:CoeventDynamics}, a property which will be key in what follows. Because $\S$ is finite the filter $\f^{-1}(1)$ has a principal element, which we denote $\f^*$. This leads to a map $*:\MCE\rightarrow\EA$ with $*:\f\mapsto\f^*$, which can be extended to a involution on $\EA\times\MCE$ by defining $*:\EA\rightarrow\MCE$ with $*:A\mapsto A^*$ where for all $B\in\EA$,
\beq\label{eq:multiplicative coevent valuation}
A^*(B) = \left\{\begin{array}{cc} 1 & \text{if} \ A\subseteq B \\ 0 & \text{otherwise} \end{array}\right.
\eeq
It is easy to check that $\f$ is bijective and that $(\f^*)^*=\f$ and $(A^*)^*=A$; we can therefore think of $*$ as a duality.

\section{Coevents as Beables}\label{sec:coevents as beables}

\subsection{The General Construction}\label{subsec:coevents as beables general appraoch}

As stated above our aim is to examine the logical implications of interpreting coevents as beables. If we are to take a set of coevents seriously as our ontology then we must reconfigure the construction of our physical theories and reasoning accordingly. Taking coevents as the `genuine' beables and relegating histories to the category of emergent constructs makes it natural to formulate theories and interpretations thereof in terms of coevents rather than histories. We might for example seek a theory of the form $(\V,\EV,\Pi)$, where $\Pi$ is a generalised measure\footnote{A generalised measure is not necessarily classical or quantum \cite{Sorkin:1994dt}} on $\EV$. Ideally this shift to the `genuine' beables would lead to a simpler dynamics, perhaps even a probability measure; however efforts to construct such a probability measure have encountered obstructions \cite{Yousef:CoeventDynamics}.

In this paper we will focus not on the dynamics of coevents but on their associated logical framework. Indeed if $\V$ is truly to be regarded as the ontology then it would perhaps be more accurate to think of events in $\EA$ as valuations on $\EV$. To avoid confusion we will distinguish between the `original' \emph{histories event algebra} $\EA$, whose elements we will call \emph{histories events}, and the \emph{valuation event algebra} $\EV$ whose elements we will call \emph{valuation events}. Notice that $\EA$ and $\EV$ are both Boolean lattices under the ordering defined by inclusion. To formulate a logical framework for valuation events we begin by considering the following points,
\begin{enumerate}
\item{In the coevent literature \cite{Sorkin:coeventsI,Sorkin:coeventsII,Sorkin:2010,Yousef:CoeventDynamics,Yousef:KSP,Yousef:thesis} no attempt is made to apply `non-standard' reasoning to coevents. We could say that `standard meta-logic' is applied to coevents.}
\item{To formally construct a logical framework for coevents we must carefully `move' our reasoning from the realm of `meta-logic' to that of an `object logic'; a formal mathematical structure (which we will in turn describe using our meta-language and meta-logic). }
\item{Inspired by our treatment of classical histories theories (section \ref{subsubsec:classical ontology}) we \emph{choose} to implement `standard meta-logic' as a Boolean lattice of propositions associated with homomorphic truth valuations from the whole space to $\Z2$. Following section \ref{subsubsec:classical ontology} we expect our propositional lattice to correspond to the event algebra ordered by inclusion.}
\end{enumerate}
This leads us to use $(\EV,Hom(\EV,\Z2),\Z2)$ as the logical framework for our new beables. We will use $f$ to denote a general member of $Hom(\EV,\Z2)$, and $\fp\in Hom(\EV,\Z2)$ to denote the unique map taking $\phi\in\V$ to $1$.

However, even if we identify $\EV$ as the `genuine' event algebra, it is the events in $\EA$ to which we have experimental access, and around which we have developed dynamical theories. Taking coevents as the beables suggests that the histories events are constructs built of valuation events; we therefore seek to represent a histories event $A\in\EA$ with a valuation event. In other words we want a map,
\beq
\t:\EA\rightarrow\EV.
\eeq
The key to defining $\t$ is a shift in focus from the unevaluated proposition `$A$'($\in\EA$) to the evaluated proposition `$\phi(A)=1$', leading us to interpret `$A$ is true' as meaning `the true valuation maps $A$ to $1$'. To facilitate this we define,
\bea
\t(A) &=& \{\phi\in\V \ | \ \phi(A)=1\} \nn \\
&=& supp(A),
\eea
where $supp(A)$ denotes the support of $A$ considered as a map on $\EV$,
\bea
A:\EV&\rightarrow&\Z2 \nn \\
A(\phi)&=&\phi(A).
\eea
Now if we do not know which valuation is `real', but do know that a particular histories event $A\in\EA$ is true, then we can assert that $\t(A)$ is true, yielding a translation of histories events into valuation events,
\beq
\phi(A)=1 \Leftrightarrow \fp(A)=1.
\eeq

Notice that when $\t$ is injective we can define two order structures on $\t(\EA)=\{\t(A) \ | \ A\in\EA\}$.  We have the order `pushed forward' from $\EA$,
\beq
\t(A)\leqt\t(B)\Leftrightarrow A\leq B,
\eeq
and the order inherited from $\EV$, which is defined by inclusion,
\beq
\t(A)\leqs\t(B)\Leftrightarrow \t(A)\subseteq\t(B).
\eeq
If $\t$ is not injective $\leqs$ is unaffected, however we will have to check that $\leqt$ is well defined. As we shall see below the relationship between these two orders is a means of expressing the properties of whichever coevent scheme we are considering. We will explore the realisation of these ideas below by considering the concrete example of multiplicative coevents.

\subsection{Multiplicative Coevents}\label{subsec:mult coevents as beables}

\subsubsection{Constructing the Map $\t$}

To specify $\t$ we must first examine the order structure of multiplicative coevents more closely. Given a histories theory $\H$ the set of all multiplicative coevents is
\beq
\MCE=\{A^* \ | \ A\in\EA\}.
\eeq
As we have seen in section \ref{subsubsec:multiplicative scheme} the map $*$ is an involution on $\EA\times\EA^*$ and can thus be thought of as a duality. This leads us to define an `inverse' ordering on $\MCE$,
\beq
A^*\leq B^* \Leftrightarrow A\geq B.
\eeq

We now introduce the maps $\uparrow$ (and $\downarrow$) sending a poset $P$ to $2^P$, such that each element is sent to the set of elements greater than (less than) or equal to it. Thus for example we have,
\bea
\uparrow:P &\rightarrow& 2^{P} \nn \\
\uparrow(p) &=& \{q\in P \ | \ q\geq p\}.
\eea
We call the sets $\uparrow(p)$ \emph{filters}, and will denote the space of such filters in $\EA$ by $\F\subset {2^{\EA}}$. Similarly we will denote the space of filters in $\MCE$ by $\Fs\subset 2^{\MCE}$. Note that because $*$ is order-reversing we have,
\bea
(\uparrow(A))^*&=&\downarrow(A^*) \nn \\
(\downarrow(A))^*&=&\uparrow(A^*).
\eea
Then using (\ref{eq:multiplicative coevent valuation}) it is easy to see that,
\bea
supp(A) &=& \{B^* \ | \ B\subseteq A\} \nn \\
&=& \{B^* \ | \ B\leq A\} \nn \\
&=& (\downarrow(A))^* \nn \\
&=& \uparrow(A^*).
\eea
This leads us to,
\bea
\t:\EA&\rightarrow&\Fs \nn \\
\t(A) &=& \uparrow(A^*).
\eea
It is easy to check that $\t$ is injective.

\subsubsection{The Order Structure of $\t(\EA)$}

We have established $\Fs=\t(\EA)$ as the `genuine' set of valuation events which we are `actually' accessing when we discuss histories events; we now turn to the order structure of this set. We have two `natural' orders on $\Fs$, the first pushed forward from $\EA$ and the second inherited from $\EM$.
\begin{enumerate}

\item{Since $\t$ is injective onto $\Fs$ we can push forward the order, and thus lattice, structure of $\EA$,
\beq
\t(A)\leqt\t(B) \Leftrightarrow A\leq B
\eeq
This makes $\t$ a lattice isomorphism onto $\Fs$ using the order $\leqt$ and the related operations $\wedget,\veet$.}

\item{It is easy to see that $2^{\MCE}$ is a Boolean lattice using the standard inclusion ordering ($\a\leqs\b\Leftrightarrow\a\subseteq\b$). Then $\Fs\subset 2^{\MCE}$ inherits this ordering. Note that in general a subposet need not be a sublattice, for example the subposet may not contain meets and joins defined using the ambient space ordering.   }

\end{enumerate}

We can now compare these two orderings. We have,
\bea
\t(A)\leqt\t(B) &\Leftrightarrow& A\leq B \nn \\
&\Leftrightarrow& A^*\geq B^* \nn \\
&\Leftrightarrow& \uparrow(A^*)\subseteq\uparrow(B^*) \nn \\
&\Leftrightarrow& \t(A)\leqs\t(B) \label{eq:mult coevents AND algebra}
\eea
so that $\leqt = \leqs|_{\Fs}$; the ordering pushed forward from the histories event algebra is the same as the order inherited from the coevent algebra. We can then think of $\wedget$ and $\veet$ as being defined by taking sup's and inf's within $\Fs$ using the inherited order. However there is no guarantee that the inf's and sup's taken within $\Fs$ are the same as those taken within $\EM$, since for example an inf within $\Fs$ is the greatest among elements contained in $\Fs$ whereas an inf in $\EM$ is free to consider elements not in $\Fs$.

We must check whether meets and joins defined by $\leqs$ in $\EM$ are equivalent to those defined by its restriction to $\Fs$. We begin with meets,
\bea
\t(A)\wedges\t(B) &=& \t(A)\cap\t(B) \nn \\
&=& \uparrow(A^*)\cap\uparrow(B^*) \nn \\
&=& \uparrow(A^*\vee B^*) \nn \\
&=& \uparrow((A\wedge B)^*) \nn \\
&=& \t(A\wedge B) \nn \\
&=& \t(A)\wedget\t(B)
\eea
by the definition of $\wedget$. However turning to joins we have,
\beq
\uparrow(A^*)\cup\uparrow(B^*)\neq\uparrow(A^*\wedge B^*)
\eeq
and thus
\beq
\t(A)\vees\t(B)\neq\t(A)\veet\t(B).
\eeq
In fact for general $A,B\in\EA$,
\beq
\t(A)\vees\t(B)=\t(A)\cup\t(B)\nsubseteq\Fs.
\eeq
Thus $\leqt$ and $\leqs$ agree on meets but not on joins. In terms on order structure this is essentially because the intersection of two filters is a filter whereas the union of two filters is not a filter.

Another way of looking at this would be to think of $\Fs$ as being closed under $\leqs$ meets but not joins. We can of course complete $\Fs$ by taking its closure under $\leqs$ joins (and meets), which we will call $\U\subset$ ${\EM}$, or its closure under $\leqs$ joins and set complements (and meets), which we will call ${\cal{C}}$. Looking forward to section \ref{subsec:topos for mult coevents} we note that any $\a\in\U$ is an upper set and that $\U$ itself is a Heyting algebra (since it is a \emph{locale}), but not in general a Boolean algebra, whereas ${\cal{C}}$ is Boolean.

\subsubsection{Truth Functions}

Taking valuations seriously as beables we have shifted from histories events in $\EA$ to valuation events in $\EM$, and have interpreted the elements of $\EA$ in terms of $\EM$ by use of $\t$. Further, we have seen that we can extend $\t(\EA)$ to $\U$ (or ${\cal{C}}$) if we wish to use joins (or complements) on the statements in $\t(\EA)$ using the `genuine' ordering in $\EM$ as opposed to the ordering pushed forward from $\EA$. Though this may at first appear relatively involved, it is a necessary step in the formalisation of statements that are quite typical concerning coevents. We are assuming that there is a single `real' coevent\footnote{We take the assumption of homomorphic truth valuation functions as equivalent to the assumption that exactly one coevent is `real'.}, however we do not know which element of $\MCE$ (or $\M$) is the `real' one. Then to say that `the real coevent maps $A\in\EA$ and to $1$ or it maps $B\in\EA$ to $1$' requires the use of $\U$. It is perhaps more clear if we turn this around; histories events $A\in\EA$ are used to give us information which narrows down the possibilities of which coevent is real, so that $\phi(A)=1$ means that $\fp(\t(A))=1$, in other words `the real coevent is in $\t(A)$'. But what if we have information about more than one event? As we will see below, though we can make statements such as `$A$ is true and $B$ is true' using only $\t(\EA)$, to make the statement `$A$ is true or $B$ is true' we will need $\U$.

More rigorously, the statement `$A$ is true and $B$ is true' translates to $\phi(A)\curlywedge\phi(B)=1$, whereas `$A$ is true or $B$ is true' translates to $\phi(A)\curlyvee\phi(B)=1$. We begin with the first statement, recalling that $\fp$ is a homomorphism and using (\ref{eq:mult coevents AND algebra}) we have,
\bea
\phi(A)\curlywedge\phi(B) &=& \fp(\t(A))\curlywedge\fp(\t(B)) \nn \\
&=&\fp(\t(A)\wedge\t(B)) \nn \\
&=&\fp(\t(A\wedge B)) \nn \\
&=&\phi(A\wedge B). \label{eq:mult coevents AND evaluation}
\eea
Our algebraic structures thus lead us back to the defining property of multiplicative coevents. Turning to the second statement we have,
\bea
\phi(A)\curlyvee\phi(B) &=& \fp(\t(A))\curlyvee\fp(\t(B)) \nn \\
&=&\fp(\t(A)\vee\t(B)) \nn \\
&\neq&\fp(\t(A\vee B)) \nn \\
&=&\phi(A\vee B). \label{eq:mult coevents OR evaluation}
\eea
This embodies the anhomomorphism of multiplicative coevents, and makes it clear that this anhomomorphism directly reflects the difference between the lattice structures induced by the orders $\leqt$ and $\leqs$ on $\t(\EA)$. If we focused only on $\EA$ we might conclude that intuitively multiplicative coevents were consistent with the `AND' (meet) but not the `OR' (join) logical operator; however since we are taking $\EM$ as the `genuine' event algebra we are able to use $(\t(A)\vee\t(B))\in\U\subseteq$ ${\EM}$ as a formal proposition. As mentioned above, propositions of this form are already used in the analysis of coevents, we have simply moved these propositions from the realm of `meta-logic' into our formal `object-logic'.

\subsubsection{The Multiplicative Scheme}\label{subsec:mult scheme as beables}

Let $\H$ be a histories theory, if we have full knowledge of the dynamics $\mu$ then instead of $\MCE$ we might use $\M$ as our ontology. Note that $\M\subset\MCE$, so that $\EMS$ is a Boolean sublattice of $\EM$. As before we define,
\bea
\t:\EA&\rightarrow&\EMS \nn \\
\t(A) &=& supp(A) = \{\phi\in\M \ | \ \phi(A)=1\}.
\eea
Then unlike the case for general multiplicative coevents $\t$ will not in general be injective. However because every $supp(\phi)$ is a filter in $\EA$ for every $\phi\in\M$, for $A,B\in\EA$ we have
\beq
\phi(A)=1, \ A\leq B \Rightarrow \phi(B)=1.
\eeq
This implies,
\beq
supp(A)\subseteq supp(B)
\eeq
which means,
\beq
A\leq B \Rightarrow \t(A)\leqs \t(B),
\eeq
and thus that the `push forward' order is well defined, $\leqt=\leqs|_{\EMS}$. Then from here on the multiplicative scheme will behave in the same way as multiplicative coevents, yielding the same results.

\subsubsection{General Coevents}\label{subsec:general coevents as beables}

A natural question is how the above results generalise to arbitrary coevent schemes. A general coevent is a map $\phi:\EA\rightarrow\Z2$, and a general coevent scheme is an arbitrary set of coevents,
\beq
\V\subseteq\{\phi \ | \ \phi:\EA\rightarrow\Z2\}.
\eeq
As before we can define our map,
\bea
\t:\EA&\rightarrow&\EV \nn \\
\t(A) &=& supp(A).
\eea
However since $\V$ is an arbitrary set of coevents we do not know if $\t$ is injective, or more generally whether the ordering $\leqt$ is well defined. Further, even if we restrict to the case in which $\leqt$ is well defined, we can not in general relate the orders $\leqt$ and $\leqs$, leaving our analysis somewhat sterile. We can of course note that $\t(\EA)\subset\EV$ and formally construct the truth functions $\fp$, however there is not much else of interest we can say in general using the above approach.

\section{The Applicability of Topos Methods}\label{sec:topos}

\subsection{Opening Comments}

The map $\t$ defined above takes a histories event $A\in\EA$ and returns the set of coevents which send that event to `true'. Intuitively, we could think of $\t(A)$ as representing the `location' in which $A$ is true. This is reminiscent of the topos interpretation of truth valuations, which can be thought of as returning a location in which a proposition might be true\footnote{A location at which it is well defined to state that the proposition in question is true.}. This suggests the applicability of topos methods; we shall very briefly ask whether and in what manner the extensive machinery of topos theory can be brought to bear on our coevent constructions, we shall leave the actual exploration and interpretation of topos constructions for coevents to further research.

We follow the application of topos theory to physics discussed by Isham, Butterfield, D\"{o}ring, Flori and others \cite{Isham:1997,Isham:toposI,Isham:toposII,Isham:toposIII,Isham:toposIV,Butterfield:2003,Flori:2008fb,Isham:1999kb}, in particular Isham's early application of topos methods to the histories approach as described in \cite{Isham:1997}; this analysis giving us an application of topos methods to structures similar to the ones we are considering. Isham's later work with D\"{o}ring \cite{Isham:toposI,Isham:toposII,Isham:toposIII,Isham:toposIV} expands on these initial ideas to propose a more comprehensive and far reaching framework whose application to our problems is beyond the scope of this paper, though it may well be grounds for fruitful future research.

\subsection{The Topos of Varying Sets}

\subsubsection{Varying Sets}

We will briefly outline the varying set structure used in \cite{Isham:1997}. For simplicity and accessibility we will suppress much of the actual topos theoretic intuition and follow the focus of this paper by concentrating on order structure; a fuller account is of course to be found in \cite{Isham:1997}.

Our central object is $\vP_Q$, the \emph{constant varying set over P}, which we can perhaps think of as a poset $P$ (which we will assume to be finite) foliated with copies of a set $Q$.
\bea
\vP_Q:P & \rightarrow & 2^{Q} \nn \\
\vP_Q(p) &=& Q \label{eq:constant varying set}
\eea
We call a map $\sP:P \rightarrow 2^{Q}$ a \emph{subobject} of $\vP_Q$ if,
\beq\label{eq:subobject}
p\leq q \Rightarrow \sP(p)\subseteq\sP(q).
\eeq

Constant varying sets (and their subobjects) are \emph{objects} within the topos $\mathbf{Sets^{P}}$, which more generally contains \emph{varying sets over P}. We can think of a varying set $X$ over a poset $P$ as the association of a set $X(p)$ with each element of $p\in P$, and the association with every pair $p\leq q$ of a map $X_{pq}:X(p)\rightarrow X(q)$ satisfying the conditions,
\begin{enumerate}
\item{$X_{pp}$ is the identity map.}
\item{$p\leq q\leq r \Rightarrow X_{pq}=X_{rq}\circ X_{pr}$.}
\end{enumerate}
Thus setting $X(p)=Q$ $\forall p\in P$ and defining the map $X_{pq}$ to be inclusion we get $X=\vP_Q$.

Now let $p\in P$ be an element of a poset $P$, we will denote by $U(p)\subset 2^P$ the set of all upper sets above $p$; then given $A\in 2^P$ we have $A\in U(p)$ iff,
\begin{enumerate}
\item{$r\in A\Rightarrow r\geq p$}.
\item{$r\in A, \ r'\geq r \Rightarrow r'\in A$}.
\end{enumerate}
Upper sets are the realisation in this context of the more general topos concept of \emph{sieves} \cite{Isham:1997}. We will denote the union of all upper sets in $P$ by $U(P)$. 

\subsubsection{The Sub-Object Classifier}

We are now in a position to introduce the \emph{sub-object classifier} $\sO$, a key object in topos theory which can be thought of as the space of truth values for the topos in question. In our case $\sO$ will be the object which associates $U(p)$ with every $p\in P$, $\sO(p)=U(p)$, equipped with the natural map $\sO_{pq}:\sO(p)\rightarrow\sO(q)$ defined by,
\beq
\sO_{pq}(A)=A\cap\uparrow (q).
\eeq

Now if our poset contains only one element, $P=\{p\}$, then the constant varying set $\vP$ is simply the set $Q$ associated to the point $p$, and a subobject $A$ is simply a subset $A\subseteq Q$ (associated with $p$). Clearly this structure is analogous to standard set theory. Our subobject classifier in this case associates to $p$ the two upper sets $\emptyset$ and $\{p\}$, which we can label $0$ and $1$ respectively; it is easy to see that the natural Heyting algebra\footnote{See \cite{Isham:1997} for further details.} of $\sO$ is then equivalent to the Boolean algebra $\Z2$. Following the standard set theory practise we can construct \emph{characteristic maps} for each subset $A\subseteq Q$,
\bea
\chi^A:Q \ (=\vP(p))&\rightarrow& \Z2 \ (=\sO(p)) \nn \\
\chi^A(x) &=& \left\{\begin{array}{cc} 1 & x\in A \\ 0 & x\not\in A \end{array}\right.
\eea
The map $\chi^A$ can be thought of as evaluating the proposition $x\in A$ and returning `true' or `false'. However, our varying set structure allows us to reinterpret $\chi$; noting that the subset $A$ corresponds to a subobject $A(p)$ and recalling that in our subobject classifier $0=\emptyset$ and $1=\{p\}$, we could equivalently write,
\beq
\chi^A(x) = \{q\in P \ | \ x\in A(q) \}, \nn
\eeq
so that $\chi$ picks out the location at which $x\in A$ is true. Then the identification of $\emptyset$ with $0$ and $\{p\}$ with $1$ allows us to think of these locations as $\Z2$ truth values. Though this may seem an unnecessary complication when $P$ contains only one element, we can use this idea to generalise to,
\beq
\chi^A:\vP \rightarrow \sO,
\eeq
where the map acts on each of the sets associated with the elements of $P$,
\bea
\chi^A_p:\vP(p)&\rightarrow& \sO(p) \nn \\
\chi^A_p(x) &=& \{q\geq p \ | \ x\in A(q) \}. \label{eq:varying set characteristic map}
\eea
Intuitively we have shifted from asking `is $x$ in $A$?' to `(given $p$) at what locations (above $p$) is $x$ in $A$?'\footnote{Or perhaps `under what conditions is $x$ in $A$?'.}.

\subsection{Topos for multiplicative coevents}\label{subsec:topos for mult coevents}

We now ask whether and how the above mechanism might apply to multiplicative coevents. As we are taking coevents as our beables we want `truth locations' to be sets of coevents. This suggests that our `base' poset should be made up of multiplicative coevents; $\MCE$ is the natural choice. This places us within the topos $\mathbf{Sets^{\MCE}}$, then following \cite{Isham:1997} we will single out a constant varying set in this topos and represent our physical theory as a subobject of this constant varying set. Since the propositions accessible to us are histories events we are led to,
\bea
\vM:\MCE&\rightarrow& 2^{\EA} \nn \\
\vM(\phi) &=& \EA.
\eea

We turned to topos theory because $\t(A)$ intuitively seems like a `truth location'; in particular for multiplicative coevents $\t(A)$ is a filter and thus an upper set in $\MCE$, matching the structure of `truth locations' in $\mathbf{Sets^{\MCE}}$. We then seek a subobject of $\vM$ intuitively corresponding to $\t(A)$; the most obvious choice is,
\bea
\vsupp:\MCE&\rightarrow& 2^{\EA} \nn \\
\vsupp(\phi) &=& supp(\phi).
\eea
Now,
\bea
A^*\leq B^* &\Rightarrow& A\geq B \nn \\
&\Rightarrow& \uparrow(A)\subseteq\uparrow(B) \nn \\
&\Rightarrow& supp(A^*)\subseteq supp(B^*), \label{eq:vsupp is a subobject}
\eea
so that $\vsupp$ is a genuine subobject of $\vM$. Then our characteristic map is,
\beq
\chi^{\vsupp}_{B^*} = \{C^*\geq B^* \ | \ A\in supp(C^*) \}.
\eeq
Then noting the duality between a coevent acting on an event and an event acting on a coevent we see that $A\in supp(C^*)\Leftrightarrow C^*\in supp(A) = \t(A)$, so that,
\bea
\chi^{\vsupp}_{B^*} &=& \{C^*\geq B^* \ | \ C^*\in\t(A) \} \nn \\
&=& \uparrow(B^*)\cap\uparrow(A^*) \nn \\
&=& \t(A\wedge B).
\eea
Alternately we could write this as,
\beq
\chi^{\vsupp}_{\phi}(A)=\t(A\wedge\phi^*).
\eeq
This implements the conception of `partial' or `contextual' truth \cite{Isham:1997}; since our base poset is $\MCE$ itself the coevents $\phi\in\MCE$ act as the `contexts' for our propositions $A\in\EA$.

If we continue to follow \cite{Isham:1997} we would then assemble these `local' characteristic maps in to a global element of $\sO$ which could be regarded as a `valuation' of pairs $(A,\phi)$ where $A\in\EA$ and $\phi\in\MCE$. The details are unnecessary for our analysis, so we will simply note that intuitively this valuation assigns $\chi^{\vsupp}_{\phi}(A)=\t(A\wedge\phi^*)$ to the pair $(A,\phi)$.


\subsection{Discussion}

First we note that the above construction depends on the properties of $\MCE$, in particular (\ref{eq:vsupp is a subobject}) shows that $\vsupp$ is a subobject of $\vM$ by using properties of multiplicative coevents which do not generalise to arbitrary coevents. Thus we can not generalise the above construction to arbitrary coevents schemes.

Secondly we consider the multiplicative scheme. If the dynamics $\mu$ of a histories theory $\H$ is known then we might prefer to use the multiplicative scheme $\M$ rather than the full set of multiplicative coevents $\MCE$. For the above this could mean replacing $\MCE$ with $\M$ as the base poset of our topos; which would become $\mathbf{Sets^{\M}}$. Since $\M\subset\MCE$ it is easy to check that the above analysis holds; however primitivity ensures that $\M$ is an anti-chain so the structures based on the constant varying set $\triangle\M$ may not be particularly illuminating.

These caveats made we conclude that we can formulate multiplicative coevents in a manner that allows the application of topos theory as used in \cite{Isham:1997}, though this construction does not generalise to arbitrary coevent schemes and may not be interesting for the multiplicative scheme. This formulation enables us to utilise new methods in understanding coevents, for example by expressing multiplicative coevents within the comprehensive framework introduced by Isham and D\"{o}ring in \cite{Isham:toposI}. Further, such constructions provide a context in which multiplicative coevents might be compared to other approaches to `quantum logic'.

\section{Conclusion}

Our aim was to develop a logical framework for coevents, taking them `seriously' as beables, using this to compare  propositions concerning coevents with propositions concerning histories. Outlining our general approach in section \ref{subsec:coevents as beables general appraoch}, we identified the lattice of coevent propositions as $\EV$ and introduced the map $\t$ which allows us to represent histories propositions in $\EA$ as elements of $\t(\EA)\subset\EV$; we found that a comparison of the two logical structures then reduced to a comparison of the lattice structure pushed forward from $\EA$ to $\t(\EA)$ and the lattice structure inherited by $\t(\EA)$ from $\EV$. Applying these ideas to multiplicative coevents in section \ref{subsec:mult coevents as beables} we found that the map $\t$ took the elegant form $\t(A)=\uparrow(A^*)$, leading to the equivalence of the pushed forward and inherited orderings on the image of $\t$, $\leqt=\leqs |_{\Fs}$. However this did not lead to the two lattice structures being equivalent, in fact we found that $\wedget=\wedges |_{\Fs}$ but $\veet\neq\vees |_{\Fs}$ reflecting the anhomomorphic nature of multiplicative coevents, as became more clear once we considered the truth valuations $\fp$. We further considered how this analysis might apply to the multiplicative scheme and more general coevent schemes in sections \ref{subsec:mult scheme as beables} and \ref{subsec:general coevents as beables}. In section \ref{sec:topos} we noticed that the map $\t$ intuitively resembled topos notions of truth valuation, and used this as a handle to suggest a means by which topos theory (as applied by Isham) might be applied to multiplicative coevents; a result which we hope opens the way for future research.

\newpage
\bibliography{Bib}
\bibliographystyle{plain}

\end{document}